\documentclass[12pt]{iopart}

\usepackage{iopams,graphicx,amssymb}

\begin{document}

\title[Anomalous scaling and large-scale anisotropy in
magnetohydrodynamic turbulence]
{Anomalous scaling and large-scale anisotropy in
magnetohydrodynamic turbulence: Two-loop renormalization-group
analysis of the Kazantsev--Kraichnan kinematic model}

\author{N.\,V.~Antonov$^{1}$ and N.\,M. Gulitskiy$^{1,2}$}

\address{$^{1}$ {\it Department of Theoretical Physics,
St~Petersburg
University, Uljanovskaja~1, St~Petersburg---Petrodvorez, 198904 Russia} \\
$^{2}$ {\it D.\,I.~Mendeleev Institute of Metrology, Moskovsky~pr.~19,
St~Petersburg, 190005~Russia}}

%\maketitle

\begin{abstract}
The field theoretic renormalization group and operator product expansion are
applied to the  Kazantsev--Kraichnan kinematic model for the
magnetohydrodynamic turbulence. The anomalous scaling emerges as a
consequence of the existence of certain composite fields
(``operators'') with {\it negative} dimensions. The anomalous
exponents for the correlation functions of arbitrary order are
calculated in the two-loop approximation (second order of the
renormalization-group expansion), including the anisotropic
sectors. The anomalous scaling and the hierarchy of anisotropic
contributions become stronger due to those second-order contributions.
\end{abstract}


\maketitle

Much attention has been attracted to the problem of intermittency and
anomalous scaling in developed magnetohydrodynamic (MHD) turbulence;
see e.g. \cite{GM}--\cite{SW1} and references therein.
It has long been realized that in the so-called Alfv\'enic regime, the
MHD turbulence demonstrates the behavior, analogous to that of the
ordinary fully developed fluid turbulence: cascades of energy from
the energy-containing range towards smaller scales, where the dissipation
effects dominate the dynamics, and self-similar (scaling) behavior of
the energy spectra in the intermediate (inertial) range.
However, intermittent character of the fluctuations in the MHD
turbulence is much stronger pronounced than that in ordinary fluids.

The solar wind provides an ideal wind tunnel in which various models
and approaches to MHD turbulence can be tested \cite{SW2}--\cite{SW6}.
In solar flares, the highly energetic and anisotropic large-scale
events coexist with small-scale coherent (singular) structures, finally
responsible for the dissipation. Thus modeling the way in which the energy
is transferred along the spectra and then dissipated is a difficult task.
The intermittency modifies the scaling behavior of the higher-order
spectra, leading to anomalous scaling, characterized by infinite families
of independent exponents.

A simplified description of that situation was proposed in \cite{E}: the
large-scale field $B^{0}_{i} = n_{i} B^{0}$ dominates the dynamics in the
distinguished direction
${\bf n}$, while the activity in the perpendicular plane can be
approximated as nearly two-dimensional. This picture allows for reliable
numerical simulations which show that the turbulent fluctuations tend to
organize in rare coherent structures separated by narrow current sheets.
On the other hand, the observations and simulations show that the scaling
behavior in the solar wind is more similar to the anomalous scaling of
fully developed hydrodynamic turbulence, rather than to simple
Iroshnikov--Kraichnan scaling suggested by two-dimensional picture with
the inverse energy cascade; see e.g. the discussion in \cite{SW2}. Thus
further analysis of more realistic three-dimensional models is desirable.

In this paper, we study the inertial-range scaling behavior within the
framework of a simplified three-dimensional model, known as the
Kazantsev--Kraichnan kinematic model \cite{KA68}, in which the magnetic
field is passive (no feedback on the velocity), while the velocity field
is modeled by a Gaussian ensemble with prescribed statistics; see also
\cite{V96}--\cite{DV}.

In spite of their relative simplicity, the models of passive scalar fields
advected by such ``synthetic'' velocity ensembles proved to be
very interesting
because of the insight they offer into the origin of intermittency and
anomalous scaling in the real fluid turbulence on the whole: they reproduce
many of the anomalous features of genuine turbulent mass or heat transport
observed in experiments; see the review paper \cite{FGV} and the
literature cited therein.

Owing to the presence of a new stretching term in the dynamic equation,
the behavior of the passive vector field appears much richer
than that of the scalar field: ``...there is considerably more life in
the large-scale transport of vector quantities''
(p. 232 of Ref. \cite{Legacy}). Indeed, passive vector fields reveal
anomalous scaling already on the level of the pair correlation function
\cite{V96,RK97}. They also develop interesting large-scale instabilities
that can be interpreted as manifestation of the dynamo effect
\cite{KA68,V96,DV}.

In the presence of a constant background field $B^{0}_{i}$, the dynamic
equation for the fluctuating part $\theta_{i}=\theta_{i}(t,{\bf x})$
of the full magnetic field
$B_{i} = B^{0} ( n_{i}  + \theta_{i} )$ has the form
\begin{eqnarray}
\partial _t\theta_{i} + \partial_{k} (v_{k}\theta _{i} - \theta_{k}
v_{i}) = \kappa\partial^{2} \theta_{i} + n_{k}\partial_{k} v_{i}.
\label{mhd}
\end{eqnarray}
Here $v_i=v_i(t,{\bf x})$ is the velocity field,
$\kappa=c^{2}/4\pi\sigma$ is magnetic diffusion coefficient,
$c$ is the speed of light, $\sigma$ is the conductivity,
$\partial^{2}$ is the Laplace operator; summation over repeated tensor
indices is understood. Equation (\ref{mhd}) follows from the simplest form
of Ohm's law for a moving medium ${\bf j} = \sigma ({\bf E}+
{\bf v}\times{\bf B} /c)$ and the Maxwell equations neglecting the
displacement current; see e.g. \cite{Moffat}.
The last term in the right hand side of (\ref{mhd}) maintains the steady
state of the system and is a source of the anisotropy; in principle, it can
be replaced by an artificial Gaussian noise with appropriate statistics.

In the real problem, the field $v_{i}$ satisfies the Navier--Stokes
equation with the additional Lorentz force term $\sim ({\bf B} \times
{\rm curl}\,{\bf B}) $. In the Kazantsev--Kraichnan model it obeys a
Gaussian distribution with zero mean and correlation function
(we consider below the incompressible fluid)
\begin{eqnarray}
\langle v_{i}(x) v_{j}(x')\rangle = D_{0}\, \delta(t-t')\,
\int_{k>m} \frac{d{\bf k}}{(2\pi)^{d}} \, \frac{1}{k^{d+\xi}} \,
P_{ij}({\bf k})\,  \exp [{\rm i}{\bf k}\cdot({\bf x}-{\bf x'})] ,
\label{3}
\end{eqnarray}
where $P_{ij}({\bf k}) = \delta_{ij} - k_i k_j / k^2$ is the transverse
projector (needed to ensure the incompressibility condition
$\partial_i v_i=0$), $k\equiv |{\bf k}|$, $D_{0}>0$ is an amplitude factor,
$d$ is the dimensionality of the ${\bf x}$ space.
The parameter $\xi$ can be viewed as a kind of H\"{o}lder exponent,
which measures ``roughness'' of the velocity field \cite{FGV}.
The infrared (IR) regularization is provided by the cutoff in the integral
(\ref{3}) from below at $k=m$, where $m\equiv 1/L$ is the reciprocal of
the integral scale $L$. The anomalous
exponents are independent on the precise form of the IR regularization;
the sharp cutoff is the most convenient choice from the calculational
viewpoints. The natural interval for the exponent $\xi$
is $0<\xi<2$, when the so-called effective eddy diffusivity
has a finite limit for $m\to0$. However, for the magnetic field a steady
state exists only if $\xi<\xi_{c} \le 2$. In the following, we consider the
physical case $d=3$, when $\xi_{c}=1$ \cite{V96}.

The model (\ref{mhd}), (\ref{3}) corresponds to the so-called kinetic
regime, in which the effects of the magnetic field on the velocity
statistics are neglected.
In this connection, it is worth noting that in the full-scale model
of the MHD turbulence such a regime is indeed realized
in the so-called kinetic fixed
point of the renormalization group (RG)  equations \cite{MHD1}.
Various generalizations of this model (finite correlation time,
compressibility, more general form of the nonlinear terms) were also
studied \cite{alpha}--\cite{anizo}.

The RG approach to the Kazantsev--Kraichnan model is described in
\cite{Lanotte2} in detail; here we only recall the main points. The
original stochastic problem (\ref{mhd}), (\ref{3}) is reformulated
as a certain field theoretic model. The ultraviolet divergences in the
corresponding Feynman diagrams manifest themselves as poles in $\xi$.
They are removed by multiplicative renormalization; as a byproduct of that
procedure, differential RG equations are derived for various correlation
functions. They have an IR attractive fixed point. This means that in the
IR range $\Lambda r \gg1$, where $\Lambda$ is the reciprocal of the inner
(dissipation) length, the correlation functions acquire scaling forms
with certain critical dimensions $\Delta_{F}$ of all the fields and
parameters $F$ of the model.

The most important part is played by the critical dimensions
$\Delta_{n,l}$ associated with the irreducible tensor composite fields
(``local composite operators'' in the field theoretic terminology)
built solely of the fields $\theta$ at a single space-time point
$x=(t,{\bf x})$. They have the forms
\begin{equation}
F_{n,l}\equiv \theta_{i_{1}}(x)\cdots \theta_{i_{l}}(x)\,
\left(\theta_{i}(x)\theta_{i}(x)\right)^{p} + \dots,
\label{Fnl}
\end{equation}
where $l\le n$ is the number of the free vector indices and $n=l+2p$ is
the total number of the fields $\theta$ entering into the
operator; the tensor indices and the argument $x$ of the symbol
$F_{n,l}$ are omitted. The ellipsis stands for the appropriate
subtractions involving the Kronecker delta symbols, which ensure
that the resulting expressions are traceless with respect to
contraction of any given pair of indices, for example,
$\theta_{i}\theta_{j} - \delta_{ij}(\theta_{k}\theta_{k}/d)$ and so on.

The quantities of interest are, in particular, the equal-time pair
correlation functions of the operators (\ref{Fnl}). For these, solving
the corresponding RG equations gives the following asymptotic expression
\begin{equation}
\langle  F_{n,l}(t,{\bf x})F_{k,j}(t,{\bf x}') \rangle \simeq
\left(\kappa \Lambda^{2}\right)^{-(n+k)/2} %\Lambda^{-(n+k)}
(\Lambda r)^{-\Delta_{n,l}-\Delta_{k,j}}\, \zeta_{n,l;k,j}(mr)
\label{struc}
\end{equation}
with $r=| {\bf x} - {\bf x}'|$ and certain scaling functions $\zeta(mr)$.
To simplify the notation, here and below in similar expressions we
omit the tensor indices and the labels of the scaling functions.

The last expression in (\ref{struc}) is valid for $\Lambda r \gg 1$
and arbitrary
values of $mr$. The inertial-convective range
corresponds to the additional condition that $mr\ll 1$. The forms
of the functions $\zeta(mr)$ are not determined by the RG equations
themselves; their behavior for $mr\to0$ is studied using Wilson's
operator product expansion (OPE).

According to the OPE, the equal-time product $F_{1}(x)F_{2}(x')$
of two renormalized composite operators at ${\bf x} = ({\bf x}
+ {\bf x'} )/2 = {\rm const}$ and ${\bf r} = {\bf x} - {\bf
x'}\to 0$ can be represented in the form
\begin{equation}
F_{1}(x)F_{2}(x') \simeq \sum_{F} C_{F} ({\bf r}) F(t,{\bf x}),
\label{OPE}
\end{equation}
where the functions $C_{F}$ are the Wilson coefficients, regular
in $m^{2}$, and $F$ are, in general, all possible renormalized local
composite operators allowed by symmetry. More precisely, the
operators entering the OPE are those which appear in the
corresponding Taylor expansions, and also all possible operators
that admix to them in renormalization. If these operators have
additional vector indices, they are contracted with the
corresponding indices of the coefficients $C_{F}$.

It can always be assumed that the expansion in Eq. (\ref{OPE}) is made in
operators with definite critical dimensions $\Delta_{F}$. The correlation
functions (\ref{struc}) are obtained by averaging equation of the type
(\ref{OPE}) with the weight $\exp {\cal S}$, where ${\cal S}$ is the
De~Dominicis--Janssen action functional for our
stochastic problem; the quantities $\langle F \rangle$
appear on the right hand sides. Their asymptotic behavior
for $m\to0$ is found from the corresponding RG equations and
has the form $\langle F \rangle \propto  m^{\Delta_{F}}$.

From the expansion (\ref{OPE}) we therefore
find the following asymptotic expression for the scaling function
$\zeta(mr)$ in the representation (\ref{struc}) for $mr\ll1$:
\begin{equation}
\zeta(mr) \simeq \sum_{F} A_{F}\, (mr)^{\Delta_{F}},
\label{OR}
\end{equation}
where the coefficients $A_{F}=A_{F}(mr)$ are regular in $(mr)^{2}$.

The feature specific of the models of turbulence is the existence of
composite operators with {\it negative} critical dimensions. Such
operators are termed ``dangerous,'' because their contributions to the
OPE diverge at $mr\to0$ \cite{JETP}.

The dimension of the primary field $\theta(x)$ is known exactly,
$\Delta_{\theta}=\Delta_{1,1}=0$; the dimensions of the other
operators (\ref{Fnl}) are calculated as infinite series in $\xi$:
\begin{equation}
\Delta_{n,l}= \Delta_{n,l}^{(1)}\xi + \Delta_{n,l}^{(2)}\xi^{2} +
O(\xi^3).
\label{series}
\end{equation}
The dimensions $\Delta_{2,0}$ and $\Delta_{2,2}$ can be found in a closed
form \cite{Lanotte2}
by comparison with the exact results for the pair correlation function,
obtained within the zero-mode approach in \cite{V96}--\cite{Lanotte}.
For general $n$ and $l$, the first-order term in (\ref{series}) was derived
in \cite{Lanotte2} and has the form (for $d=3$ and up to the notation):
\begin{equation}
\Delta_{n,l}^{(1)}= - n(n+3)/10 + l(l+1)/5.
\label{Dnl}
\end{equation}

We have calculated the second term in (\ref{series}), which corresponds to
the two-loop approximation of the RG. The calculation is rather cumbersome
and will be discussed elsewhere (see also \cite{SL} for a brief discussion),
and here we only present the final result:
\begin{eqnarray}
\Delta_{n,l}^{(2)} = &-& \frac{2n(n-2)}{125} -
\frac{n(n+3)}{30} +\frac{22\,l(l+1)}{375}-
\nonumber \\
&-& \frac{3(n-2)}{175}\left(-{\sqrt{3}}\pi+\frac{82}{15}\right)
\Big[2n(n-4)+3l(l+1)\Big]-
\nonumber \\
&-& \frac{19(n-2)}{350}\left(-{\sqrt{3}}\pi+\frac{1568}{285}\right)
\Big[n(n+3)-2l(l+1)\Big] .
\label{Ans}
\end{eqnarray}
In particular, for the two special cases mentioned above this gives
\begin{equation}
\Delta_{2,0} =-\xi- \xi^2/3 +O(\xi^3), \quad
\Delta_{2,2} = \xi/5 +7\xi^2/375 +O(\xi^3)
\label{exact}
\end{equation}
in agreement with the exact results of \cite{V96}--\cite{Lanotte}. This
confirms validity and mutual consistency of the zero-mode and the RG+OPE
approaches.

From the leading-order expression (\ref{Dnl}) it follows that the
dimensions (\ref{series}) satisfy the hierarchy relations:
$\Delta_{n,l}>\Delta_{n,l'}$ if $l>l'$, which are conveniently
expressed as inequalities $\partial \Delta_{n,l} / \partial l >0$.
This fact, first established in \cite{Lanotte}, has a deep physical
meaning: in the presence of large-scale anisotropy, the leading
contribution in the inertial-range behavior $mr\ll1$ of the
correlation functions like (\ref{struc}) is given by the isotropic
``shell'' ($l=0$). The corresponding anomalous exponent is the same as for
the purely isotropic case. The anisotropic contributions
give only corrections which vanish for $mr\to0$, the faster
the higher the degree of anisotropy $l$ is. This effect gives some
quantitative support for Kolmogorov's hypothesis of the local
isotropy restoration and appears rather robust, being observed for
the real fluid turbulence \cite{Hier}
and the scalar Kraichnan model \cite{A99}.

From (\ref{Ans}) it follows that the $O(\xi^2)$ term in the dimension
(\ref{series}) satisfies the same inequality,
$\partial \Delta_{n,l}^{(2)} / \partial l \simeq (2l+1)
(0.0053n+0.0482) >0$, and we conclude that the hierarchy of anisotropic
contributions survives, and even becomes stronger, when the second-order
corrections are taken into account.

For the most important case of the scalar operator (\ref{Fnl}) with
$l=0$ from (\ref{Ans}) one obtains
\begin{equation}
\Delta_{n,0}^{(2)} \simeq -0.0041 n^{3} - 0.0474 n^{2} - 0.0553 n,
\label{Ska}
\end{equation}
which is negative for all $n$. Thus the anomalous scaling
also becomes stronger pronounced when the $O(\xi^2)$ term is included.
In this respect, the magnetic model differs drastically from its scalar
counterpart, where the higher-order corrections to the $O(\xi)$ term
are positive, which eventually leads to the disappearance of the anomalous
scaling for $\xi\to2$; see the discussion in \cite{cube}.

To conclude with, we calculated the anomalous exponents in the
Kazantsev--Kraichnan kinematic dynamo model in the two-loop order
(including the anisotropic ``shells'' in the presence of large-scale
anisotropy). We found that both the
anomalous scaling and the hierarchy of anisotropic contributions become
stronger due to the second-order corrections to the leading terms.
It is interesting to see how these results are affected by compressibility,
anisotropy and the influence of the magnetic field on the velocity field
dynamics. This work remains for the future and is partly in progress.

\section*{Acknowledgments}

The authors are indebted to L. Ts. Adzhemyan, Michal Hnatich, Juha Honkonen,
Andrea Mazzino and Paolo Muratore Ginanneschi for valuable discussions.
N.M.G. thanks the Organizers of the Conference ``Mathematical Modeling and
Computational Physics'' (Stara Lesna, Slovakia, July 2011) for the
possibility to present the results of this work \cite{SL}.

In the published version of this paper [Phys. Rev. E{\bf 85} (2012)
065301(R)] there was an error in the third line of equation (\ref{Ans})
and, as a consequence, in expressions for
$\partial \Delta_{n,l}^{(2)} / \partial l$ and $\Delta_{n,0}^{(2)}$.
The corrected result is presented in [Phys. Rev. E{\bf 87} (2013) 039902(E)].
The main conclusions of the paper are not affected by that error:
the anomalous scaling and the hierarchy of anisotropic contributions
are enhanced by the two-loop correction. The authors are indebted to
Marian Jur\v{c}i\v{s}in for pointing out the error to us.

\end{document}